\documentclass[preprint2]{emulateapj}
\providecommand{\bm}[1]{\mbox{\boldmath{$#1$}}}
\usepackage[a4paper,lmargin=2.4cm,rmargin=0.6cm,bmargin=0cm,tmargin=5.5cm]{geometry}

\shorttitle{Magnetic Fields in Fully Convective Spheres}
\shortauthors{Dobler et al.}
\slugcomment{Received 2004 October 26; accepted 2005 October 7}

\newcommand{\Av}    {\mathbf{A}}
\newcommand{\Bv}    {\mathbf{B}}
\newcommand{\fv}    {\mathbf{f}}
\newcommand{\jv}    {\mathbf{j}}

\newcommand{\uv}    {\mathbf{u}}

\newcommand{\nnabla}{\bm{\nabla}}

\newcommand{\const}    {\mathrm{const}}
\newcommand{\Cool}     {\mathcal{C}}
\newcommand{\cs}       {c_{\rm s}}
\newcommand{\Heat}     {\mathcal{H}}
\newcommand{\Omegavect}{\bm{\Omega}}
\newcommand{\Sheartens}{\bm{\mathsf{S}}}

\newcommand{\EE}[1]{\times 10^{#1}}
\providecommand{\unit}[2][]{#1\,\mathrm{#2}}

\newcommand{\dd}{{\rm d} {}} 

\begin{document}

\title{Magnetic Field Generation in Fully Convective Rotating Spheres}

\author{
  Wolfgang~Dobler\altaffilmark{1,2},
  Michael~Stix\altaffilmark{1},
  and 
  Axel~Brandenburg\altaffilmark{3}
}
  \altaffiltext{1}{Kiepenheuer-Institut f{\"u}r Sonnenphysik,
    Sch{\"o}neckstra{\ss}e~6, D-79104 Freiburg, Germany}
  \altaffiltext{2}{Department of Physics and Astronomy, University of Calgary,
    2500 University Drive NW, Calgary, AB T2N 1N4, Canada,
    \email{Wolfgang.Dobler@ucalgary.ca}}
  \altaffiltext{3}{NORDITA, Blegdamsvej 17, DK-2100 Copenhagen \O, Denmark
  }


\begin{abstract}
Magnetohydrodynamic simulations of fully convective, rotating spheres with
volume heating near the center and cooling at the surface are presented.
The dynamo-generated magnetic field saturates at
equipartition field strength near the surface.
In the interior, the field is dominated by small-scale structures, but
outside the sphere by the global scale.
Azimuthal averages of the field reveal a large-scale field of smaller
amplitude also inside the star.
The internal angular velocity shows some tendency to be constant along
cylinders and is ``anti-solar'' (fastest 
at the poles and slowest at the equator).
\end{abstract}


\section{Introduction}

The Hayashi track in the Hertzsprung--Russell diagram characterizes young stars
in hydrostatic equilibrium that are fully convective.
Other fully convective stars are low-mass
main sequence stars (M dwarfs), and some cool giants.
These stars show strong magnetic activity as is evidenced
by chromospheric emission
in H$\alpha$
\citep[e.g.][]{Hawley:ActivityLowMass,HawleyEtal:ChromosphericActivity}
and by Zeeman broadening of classical T Tauri stars
\citep[e.g.][]{JohnsKrullEtal:BFieldBPTauri}.
In the latter case, the stars are generally rapidly rotating with
rotation periods of just a few days, and it is known that the
magnetic field shows strong departures from axisymmetry
\citep{JohnsKrullEtal:SpectrapolarimetryBPTauri}.
However, for less massive stars (M9 dwarfs and beyond) there is a sharp
decline in chromospheric magnetic activity
\citep[e.g.][]{GizisEtal:NewNeighbors},
which may be connected with
dust formation and the almost fully neutral photospheres
\citep{MohantyBasri:RotationActivity}.

Despite some progress in low-resolution Doppler imaging
\cite[e.g.][]{JoncourtEtal:DopplerImgV410Tau}, not much is known about the
surface differential rotation of these 
stars, and even less is known about their internal angular and meridional
velocities.
Theory suggests that the absolute differential rotation in fully
convective stars
decreases with increasing overall angular velocity due to rotational
quenching of the turbulent transport effect that causes the
differential rotation
\citep{KuekerEtal:AlphaOmegaRotation,%
  KitchatinovEtal:LambdaQuenching,%
  KuekerRuediger:TTauriDiffRot,%
  KuekerRuediger:TTauriAlpha2}.
As in the solar case, the equator is still predicted to rotate
more rapidly than the poles.
However, some observations of rapidly rotating stars support what is
sometimes referred to as ``anti-solar'' differential rotation,
where the equator spins less rapidly than the poles
\citep{BarnesEtal:RotationSpotsHKAqr,%
  KitchatinovRuediger:AntiSolar,%
  WeberEtal:AntiSolar}.
Since differential rotation enters as an important ingredient in
dynamo theory, it is important to develop self-consistent models
of the large-scale velocity field in fully convective stars.

Magnetic field generation in fully convective stars
is also interesting from a dynamo theoretical point of view.
With the realization that the magnetic field inside stars might be highly
intermittent and concentrated in thin flux tubes, the question of storing
such intermittent and strongly buoyant magnetic fields over the course
of the 11 year cycle became a growing concern
\citep[e.g.][]{Moreno-Insertis:RiseTimes}.
This led to the proposal that dynamos in convective shells (as in the case
of the Sun) might operate at or below the bottom of the convection zone.
This scenario would not be applicable to fully convective stars, because
they lack the overshoot layer where strong flux tubes could be stored.
However, it is known that the chromospheric activity does not disappear
for later spectral types, i.e.\ towards fully convective stars
\citep{Vilhu:Activity,VilhuEtal:Activity,
BergerEtal:RadioXrayHalphaObservations}.
It has therefore been claimed
\citep{DurneyEtal:Generation,%
  HawleyEtal:ActivityLowMass}
that fully convective stars lack
large-scale magnetic fields, but can still have small-scale fields
generated by non-helical near-surface turbulent dynamo processes.

Attempts to model such small-scale dynamo action
\citep{DorchLudwig:SmallscaleMdwarf} have however led to the conclusion
that the photospheric conductivities of M-dwarfs are most probably too low to
allow for local small-scale dynamo action
This would imply that the observed magnetic activity must be due to dynamo
action in deeper layers.

From a kinematic mean-field $\alpha^2$ dynamo model,
\cite{KuekerRuediger:TTauriAlpha2} predicted that rapidly rotating
(Coriolis number of 3 or larger) fully convective stars 
generate a non-axisymmetric steady magnetic field of quadrupolar symmetry
and azimuthal order $m=1$
that looks roughly like a dipole field with the dipole axis
lying in the equatorial plane.

Global models of convective dynamos are still in their infancy, even
though some tremendous progress was made some 20 years ago when
\cite{Gilman:DynamosII} and \cite{Glatzmaier:ConvDynamosII}
presented the first simulations
of dynamos in a spherical shell representing a solar-like convection zone.
These models predicted cyclic magnetic fields propagating toward the
poles, in contrast to the solar case.
The reason for this discrepancy remains a matter of debate even today,
when much higher numerical resolution is available.
Recent simulations still predict angular velocity to be roughly
constant on cylinders, although some simulations show at least a
tendency toward solar-like angular velocity contours
\citep{MieschEtal:3dSphericalI,%
  BrunToomre:TurbulentConvection}.
Recent simulations of dynamo action in spherical shells now begin to
produce useful models of global turbulent dynamos
\citep{Brun:Interaction,%
  BrunEtal:GlobalScale}.
Meanwhile, such global models have also been applied to core convection
\citep{BrowningEtal:CoreConvection} and to dynamo action in these cases
\citep{BrunEtal:CoreConvDynamo}.

In this paper, we present global dynamo simulations
in spheres using a Cartesian grid, i.e.~the sphere is embedded in a
cubic box.
This may seem to be an unnatural approach to spherical geometry, but it
has distinct practical advantages.
First, it avoids the coordinate singularity at the center when using
spherical coordinates, without invoking expensive transformations
from and to spherical harmonics.
Second, this approach has proven useful in view of computational
simplicity and numerical parallelization efficiency; it has
recently been applied by a number of groups to purely hydrodynamic
simulations
\citep{PorterEtal:SlabSpheroidal,%
  FreytagEtal:SpotsBetelgeuse,%
  WoodwardEtal:3dStellarConvection},
and attempts have already been made to model dynamo action in this
approach \citep{Dorch:BetelgeusePencil}.

\section{The Model}

\subsection{Basic setup}
\label{BasicSetup}

In our model the star is described as a spherical subregion of radius $R$
of a cubic box of size $L_{\rm box}^3$.
The gas in the box is governed by the usual equations of
magnetohydrodynamics (see below) with impenetrable boundaries on the box
faces such that the mass $M_{\rm box}$ in the box is
conserved.
If the gravitational well $\Phi(r)$ is sufficiently deep,
most of the mass $M$ of the star is concentrated near the center,
so $M\approx M_{\rm box}$.
Using a Newtonian cooling term in the energy equation,
the temperature outside the star is kept close to
the nominal surface temperature of the star, $T_{\rm surf}$.
An entropy gradient is maintained by prescribing a distributed
energy source $\Heat(r)$ at the center (here $r$ is the spherical radius).
The total luminosity is then given by
$L=4\pi\int_0^R \Heat(r) r^2\dd r$,
and corresponds to the energy produced by nuclear burning.
We recall however, that some young stars on the Hayashi track have
not ignited yet, and are sustaining their energy losses by contraction,
which results in a less localized energy source than nuclear fusion
reactions.
Although the mass distribution can change during the evolution of
our model, we have chosen to ignore self-gravity.

The model is governed by five main input parameters: mass $M$,
radius $R$, luminosity $L$, surface temperature $T_{\rm surf}$, and
average angular velocity $\Omega_0$.
We choose parameters that are typical of M dwarfs, but we limit
the degree of stratification to values that are numerically
more feasible by choosing a surface temperature that is much higher than
for real M dwarfs.
We also keep the Kelvin-Helmholtz time scale at a
much smaller multiple of the dynamical time scale than what is realistic.
As is common in deep convection simulations
\citep[e.g.,][]{ChanSofia:ConvectionIII,%
  BrandenburgEtal:RadiativeFluxConvection},
we do this by choosing a luminosity that is much larger than the stellar
value, and at the same time we keep the radiative diffusivity much larger
than in reality.
Since the Rayleigh number is, for a given Prandtl number,
inversely proportional to the square of the radiative diffusivity, a large
luminosity translates to a small Rayleigh number.
The restriction to moderate values of the Rayleigh number is a
common problem of all astrophysically meaningful convection simulations.

Our initial state is derived from a spherically symmetric,
isentropic reference model;
for details see Appendix \ref{RefModel}.
This state is perturbed by adding weak velocity and magnetic fields that
are both random.

\subsection{Equations}

In the computational domain, $-L_{\rm box}/2\leq x,y,z\leq L_{\rm box}/2$,
we solve the equations of compressible magnetohydrodynamics,
\begin{eqnarray}
  \frac{\mathrm{D}\ln\varrho}{\mathrm{D}t}
  &=& -\nnabla\cdot\uv \; , \\
  \frac{\mathrm{D}\uv}{\mathrm{D}t}
  &=& -\frac{\nnabla p}{\varrho}
      + \frac{\jv\times\Bv}{\varrho}
      + \frac{2}{\varrho} \nnabla\cdot\left(\varrho\nu\Sheartens\right)
      \nonumber\\ && \mbox{}
      - \nnabla\Phi
      - 2 \bm\Omega_0\times\uv
      + \fv_{\rm d} \; , \\
  \frac{\partial\Av}{\partial t}
  &=& \uv\times\Bv - \eta\mu_0\jv \; , \\
  \varrho T \frac{\mathrm{D}s}{\mathrm{D}t}
  &=& \Heat - \Cool
      + \nnabla\cdot(K\nnabla T)
      \nonumber\\ && \mbox{}
      + \mu_0\eta \jv^2 + 2\varrho\nu\Sheartens^2 \; ,
\end{eqnarray}
where $\varrho$ and $p$ denote mass density and pressure of the fluid,
$s$ and $T$ are specific entropy and temperature, $\uv$ is the fluid velocity,
$\nu$ the kinematic viscosity, $\Phi$ the gravity potential,
$\Omegavect_0$ the angular velocity of the reference frame,
$\fv_{\rm d}$ is an artificial damping force discussed in
Sec.~\ref{ProfileFunctions}, and
\begin{equation}
  \mathsf{S}_{ik}
  = \frac{1}{2}\left(\frac{\partial u_i}{\partial x_k}
                    +\frac{\partial u_k}{\partial x_i}
    - \frac{2}{3} \delta_{ik} \nnabla\cdot\uv\right)
\end{equation}
is the traceless rate-of-strain tensor.
The magnetic vector potential $\Av$ is related to the flux density
$\Bv=\nnabla\times\Av$ and the current density
$\jv=\nnabla\times\Bv/\mu_0$, and $\eta$
denotes the magnetic diffusivity.
Volume heating $\Heat$ and cooling $\Cool$ are described in
Sec.~\ref{ProfileFunctions} below.
The radiative conductivity $K$ is related to the thermal
diffusivity $\chi\equiv K/(c_p\varrho)$.
In the numerical calculations shown below, we assume $\chi$,
$\nu$, and $\eta$ to be constant across the whole box.
Our equation of state is that of a perfect gas with adiabatic index
$\gamma=5/3$.

For the gravity potential $\Phi(r)$ we choose a Pad{\'e} approximation
obtained from our isentropic reference model (see Appendix \ref{RefModel}),
\begin{equation}
  \Phi(r)
  =-\frac{GM}{R}
    \frac{a_0 + a_2 r'^2 + a_3 r'^3}
         {1 + b_2 r'^2 + b_3 r'^3 + a_3 r'^4} \; ,
\end{equation}
with $r' \equiv r/R$; we find that the coefficients,
$a_0=2.34$, $a_2=0.44$, $a_3=2.60$, $b_2=1.60$, $b_3=0.21$,
yield a good approximation both inside and outside the star.

Note that, while retaining the Coriolis force term,
we neglect the centrifugal force.
This is necessary for practical reasons, since together with the
luminosity our turbulent velocities $u_{\rm rms}$ are exaggerated and we thus
need far too large angular velocities in order to reach realistic
Coriolis numbers [see Eq.~(\ref{Eq-Rossby-num}) below].
It would therefore be unrealistic to include the strongly exaggerated
centrifugal force in the expression for the inertial forces.
We emphasize that this kind of restricted mechanics does not violate
the balance of angular momentum $\mathbf{L}$ in any significant manner:
The component $L_z$ parallel to the rotation axis is strictly conserved
(the centrifugal force is a central force for that axis), while the
other two components are small for a nearly axisymmetric system.

Our boundary conditions on the faces of the cubic box are impenetrable
free-slip conditions for velocity
($u_n=0$, $\partial_n \uv_{\rm tan}=\mathbf{0}$),
and normal-field conditions for the magnetic field
($\partial_n B_n=0$, $\Bv_{\rm tan}=\mathbf{0}$).

All numerical calculations were done using the \textsc{Pencil Code},%
\footnote{
  \url{http://www.nordita.dk/software/pencil-code} ---
  This code uses the Message Passing Interface (MPI) library for
  communication between processors and runs quite efficiently on clusters.
  Toroidal averages, spectra, and other diagnostics can be calculated
  during the run, which avoids extensive post-processing of the data.
}
a high-order centered
finite-dif\-fer\-ence
code (sixth order in space and third
order in time) for solving the compressible hydromagnetic equations.
Weak shock-capturing viscosities were used to cope with localized,
transient events of supersonic flow. 
A high-order upwind scheme is used for the advection operators for density
and entropy; see Appendix \ref{S-upwind}.

\subsection{Profile functions}
\label{ProfileFunctions}

As outlined in Sec.~\ref{BasicSetup}, the thermal structure of the star
is maintained by prescribing a 
certain distribution of heating and cooling functions
inside and outside the star, respectively.
The profile functions depend on spherical radius
$r\equiv(x^2{+}y^2{+}z^2)^{1/2}$.
In the exterior, $r>R$, we add a velocity damping term
in order to prevent excessive velocities outside the star,
which are not directly relevant to the dynamics inside the star.


\begin{figure}[tp]
  \plotone{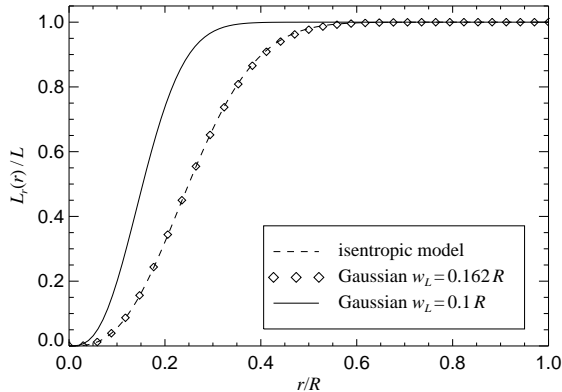}
  \caption{Comparison of luminosity function
    $L_r(r)\equiv \int_0^r \mathcal{H}(r')\,4\pi r'^2 \,dr'$ according to
    Eq.~(\ref{Heat-rho-Temp}) (dashed line) with our Gaussian
    parameterization~(\ref{Heating-prof-Gauss}).
    The choice $w_{\rm L} = 0.162\,R$ gives an excellent fit,
    while the narrower profile would be more appropriate for a heavier
    star.
  }
  \label{Fig-Lr-r}
\end{figure}

The central parts of the sphere are heated according to a normalized
Gaussian profile,
\begin{equation} \label{Heating-prof-Gauss}
  \Heat(r) = \frac{L}{(2\pi w_L^2)^{3/2}} e^{-r^2/(2 w_L^2)} \; ,
\end{equation}
which gives an excellent fit to the heating rate calculated according to
Eq.~(\ref{Heat-rho-Temp}) for our isentropic reference model if
the width $w_L$ of the nuclear burning region is chosen as $w_L=0.162\,R$.
Fig.~\ref{Fig-Lr-r} shows a comparison of the resulting luminosity from
the two parameterizations.
Most of our simulations use that value of $w_{\rm L}$, while some runs
have been carried out with $w_{\rm L}=0.1\,R$, which would be more appropriate
for a more massive star.

For $r > R$, we apply a Newtonian cooling term of the form
\begin{equation} \label{Cool-T-Tref}
  -\Cool(r) = - \varrho c_p\, \frac{T{-}T_{\rm surf}}
                     {\tau_{\rm cool}}
              \,f_{\rm ext}(r)
\end{equation}
to keep the temperature close to the surface value $T_{\rm surf}$.
Here, $f_{\rm ext}(r)$ is a profile function that smoothly interpolates
between $0$ for $r \ll R$ and $1$ for $r \gg R$.
Our profile function is a $\tanh$ profile,
\begin{equation} \label{tanh}
  f_{\rm ext}(r)
  = \frac{1}{2}\Bigl(1+\tanh\frac{r{-}R_{\rm cool}}{w_{\rm cool}}\Bigr) \; ,
\end{equation}
where $R_{\rm cool}$ and $w_{\rm cool}$ denote the position and
width of the transition.
We have chosen $w_{\rm cool}=0.05\,R$, and $R_{\rm cool}=1.05\,R$,
i.e.~slightly larger than the stellar radius,
in order to reduce the influence of the cooling term
(\ref{Cool-T-Tref}) inside the star.
In the present model, the exterior has practically constant temperature
($=T_{\rm surf}$),
i.e.\ no attempt is made to model the hot corona of the star.
In fact, since we have to restrict ourselves to moderate
stratification, the temperature ratio between the center and the surface
of the model is less than 10.

\begin{deluxetable*}{llllllrrlllrr} 
  \tabletypesize{\scriptsize}
  \tablecaption{Summary of runs discussed in the paper.
  \label{T-sum-1}
  }
  \tablewidth{0pt}
  \tablehead{
    \colhead{Run}
        & \colhead{resol.}
                  & \colhead{$\nu$}
                               & \colhead{$\chi$}
                                            & \colhead{$\eta$}
                                                         & \colhead{$\mathcal{L}$}
                                                                  & \colhead{$\Omega_0$}
                                                                           & \colhead{$\lambda$}
                                                                                     & \colhead{$u_{\rm rms}^{\rm kin\,/\,sat}$}
                                                                                                           & \colhead{$B_{\rm rms}^{\rm sat}$}
                                                                                                                     & \colhead{$\frac{B_{\rm rms}}{u_{\rm rms}}$}
                                                                                                                              & \colhead{$\mathrm{Re}$}
                                                                                                                                      & \colhead{$\mathrm{Rm}$}
 }
  \startdata
    1a  & $128^3$ & $6\EE{-4}$ & $4\EE{-4}$ & $3\EE{-4}$ & $0.02$ &  $0.2$ & $0.017$ & $0.173$\,/\,$0.164$ & $0.020$ & $0.12$ & $273$ & $547$ \\
    1b  & $256^3$ & $4\EE{-4}$ & $3\EE{-4}$ & $2\EE{-4}$ & $0.02$ &  $0.2$ & $0.043$ & $0.184$\,/\         & $0.028$ & $0.18$ & $388$ & $775$ \\
    1c  & $256^3$ & $4\EE{-4}$ & $3\EE{-4}$ & $2\EE{-4}$ & $0.01$ &  $0.2$ &         & $\phantom{0.000}$\,/\,$0.130$
                                                                                                           & $0.023$ & $0.18$ & $325$ & $650$ \\
  \hline
    2a  & $128^3$ & $8\EE{-4}$ & $8\EE{-4}$ & $4\EE{-4}$ & $0.02$ &  $0.0$ & $0.009$ & $0.239$\,/\,$0.233$ & $0.018$ & $0.08$ & $291$ & $583$ \\
    2b  & $128^3$ & $8\EE{-4}$ & $8\EE{-4}$ & $4\EE{-4}$ & $0.02$ &  $0.5$ & $0.017$ & $0.213$\,/\,$0.185$ & $0.046$ & $0.25$ & $231$ & $463$ \\
    2c  & $128^3$ & $8\EE{-4}$ & $8\EE{-4}$ & $4\EE{-4}$ & $0.02$ &  $2.0$ & $0.021$ & $0.158$\,/\,$0.129$ & $0.068$ & $0.53$ & $161$ & $323$ \\
    2d  & $128^3$ & $8\EE{-4}$ & $8\EE{-4}$ & $4\EE{-4}$ & $0.02$ &  $5.0$ & $0.036$ & $0.112$\,/\,$0.087$ & $0.099$ & $1.14$ & $109$ & $218$ \\
    2e  & $128^3$ & $8\EE{-4}$ & $8\EE{-4}$ & $4\EE{-4}$ & $0.02$ & $10.0$ & $0.038$ & $0.086$\,/\,$0.068$ & $0.105$ & $1.54$ &  $85$ & $170$ \\
  \enddata
  \tablecomments{
    Diagnostic quantities listed are:
    kinematic growth rate $\lambda$ of the magnetic field;
    root-mean-square values of velocity and magnetic flux density
    $u_{\rm rms}$, $B_{\rm rms}$;
    ratio $B_{\rm rms}/u_{\rm rms}$ for the saturated state;
    kinetic Reynolds number $\mathrm{Re}$ (based on $u_{\rm rms}$);
    and magnetic Reynolds number $\mathrm{Rm}$ (based on $u_{\rm rms}$).
    For all runs shown here, the star is embedded in a cubic box of size
    $L_{\rm box} = 3\,R$.
    The gaps for Runs~1b and 1c are due to the fact that we have not
    extended Run~1b into the final saturated regime, but rather lowered
    the value of $\mathcal{L}$ and continued it as Run~1c.
  }
\end{deluxetable*} 

Outside the star, a damping term
\begin{equation}
  \mathbf{f}_{\rm d}
  = - \frac{\uv}{\tau_{\rm d}}f_{\rm ext}(r)
\end{equation}
is applied in the equation of motion to limit flow speeds to moderate values
while still allowing the exterior to
react to sudden disturbances from the stellar surface with sufficient
flexibility; the profile $f_{\rm ext}(r)$ is the same as for the cooling term,
i.e.~Eq.(\ref{tanh}) with $w_{\rm d}=0.05\,R$, and $R_{\rm d}=1.05\,R$.
By imposing fixed radial profile functions for surface cooling and velocity
damping, we suppress the possibility of irregular surfaces that would
develop, e.g.\ in red giants \citep{FreytagEtal:SpotsBetelgeuse},
but this would not apply to real M dwarfs.

\subsection{Dimensionless parameters}
\label{S-dimensions}

As mentioned in the beginning, our model is governed by the five
basic input parameters: $M$, $R$, $L$, $T_{\rm surf}$, and $\Omega_0$.
From these, we can construct three dimensionless
quantities that characterize our model:
the stratification parameter
\begin{equation}
  \xi \equiv \frac{c_{\rm s,surf}^2/\gamma}{GM/R}\; ,
\end{equation}
(where $c_{\rm s,surf}$ is the sound speed at the surface
and $G$ is Newton's gravity constant),
the dimensionless luminosity
\begin{equation}
  \mathcal{L} \equiv \frac{L}{\sqrt{G^3M^5/R^5}}\; ,
\end{equation}
and the Coriolis number (or inverse Rossby number)
\begin{equation} \label{Eq-Rossby-num}
  \mathrm{Co} = 2\Omega_0 R/u_{\rm rms} \; ,
\end{equation}
where $u_{\rm rms}$ is the root-mean-square velocity based on
a volume average over the full sphere.
The remaining degrees of freedom determine the natural units
of our system.
In particular, length will be measured in units of the stellar radius $[x]=R$,
velocity in units of $[u]=\sqrt{GM/R}$,
density in units of $[\varrho]=M/R^3$, and specific entropy in units of
$[s]=c_p$.
This implies that time is measured in units of the dynamical time
$[t]=(GM/R^3)^{-1/2}$ and
the magnetic field is measured in units of
$[B] = \sqrt{\mu_0[\varrho]}\;[u]\equiv\sqrt{\mu_0 G}\; M/R^2$.

Note that $\xi$ is the ratio of the pressure scale height at the
stellar surface to the stellar radius, so $\xi$ controls the amount
of stratification.
The second dimensionless parameter, $\mathcal{L}$, is the ratio
of acoustic (or free-fall, or dynamic) time scale to the Kelvin-Helmholtz time.
For realistic models, both $\xi$ and $\mathcal{L}$ are much less than unity.
Using typical values for an M5 dwarf ($M=0.21\,M_\odot$, $R=0.27\,R_\odot$,
$L=0.008\,L_\odot$, and $T_{\rm surf}=4000\,{\rm K})$, we find
$\xi=2.2\EE{-4}$ and $\mathcal{L}=2.4\EE{-14}$.
In the simulations presented below, we are only able to reach
values of $\xi$ and $\mathcal{L}$ that are somewhat below unity.
In all models presented here, we have $\xi=0.19$; for most models
we choose $\mathcal{L} = 0.02$ (i.e.~$\approx 10^{12}$ times higher than
for a real M5 dwarf), while
we have $\mathcal{L}=0.01$ in one of the higher resolution runs.
The necessity of exaggerated luminosities in numerical simulations of
convection was first pointed out by \cite{ChanSofia:ConvectionIII}.
For lower values of $\mathcal{L}$, yet higher numerical resolution would be
required to get sufficiently vigorous convection and dynamo action.

Other important dimensionless parameters are the kinematic and
magnetic Reynolds numbers,
\begin{equation}
  \mathrm{Re} \equiv \frac{U R}{\nu} \quad\mbox{and}\quad
  \mathrm{Rm} \equiv \frac{U R}{\eta} \; ,
\end{equation}
where $U$ is the root-mean-square (rms) velocity within the sphere of
radius $R$.
In the present simulations, $\mathrm{Re}$ and $\mathrm{Rm}$ are
in the range 100--780 (see Table~\ref{T-sum-1}).
Realistic values of the fluid and magnetic Reynolds numbers are
much larger than what can be achieved in this type of simulation.

\section{Results}

The parameters for the runs discussed and presented in this
paper are summarized in Table~\ref{T-sum-1}.
Throughout this paper, overbars denote azimuthal averages.
The rms values listed here are also averaged in time.
In Runs~1a--1c, luminosity and resolution have been varied, while in
Runs~2a--2e we have varied the angular velocity $\Omega_0$.

The simulations were typically run for about $8\,\tau_{\rm Ohm}$, where
$\tau_{\rm Ohm} \equiv R^2/(\pi^2 \eta)$ is the diffusive time scale for a
structure of wave length $2\,R$.
One exception was the higher-resolution runs 1b and 1c, which were only
run for about $0.3\,\tau_{\rm Ohm}$ each.
In all cases, the saturated state of the magnetic field was well
established (with the exception of Run~1b, which we did not run long
enough) and quasi-stationary behavior was reached.
To ensure that we are not missing any slow trends, we continued Run~2c
until $12\,\tau_{\rm Ohm}$, but found nothing new during this somewhat
prolonged saturated calculation.

For all runs listed in Table~\ref{T-sum-1} the box
size was $L_{\rm box} = 3\,R$.
To investigate the role of the boundaries of the numerical box, we did a
reference run in a larger box ($L_{\rm box} = 5\,R$) at comparable
resolution. The results were fully compatible with $L_{\rm box} = 3\,R$.


\begin{figure*}[t!]
  \plottwo{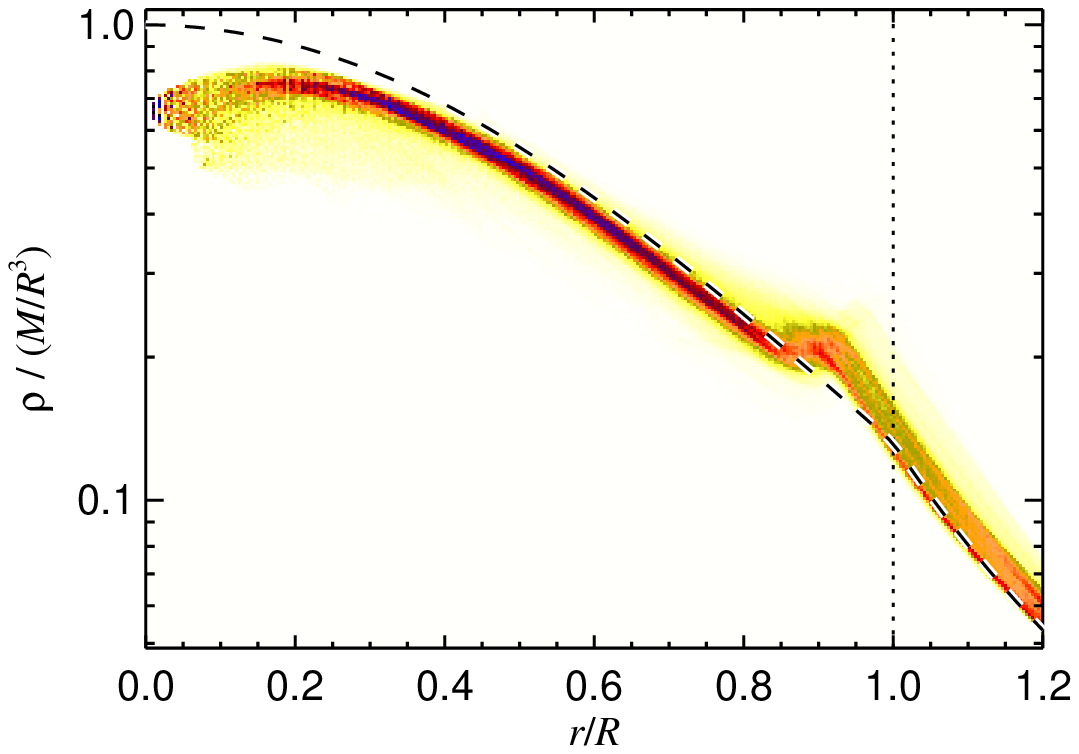}{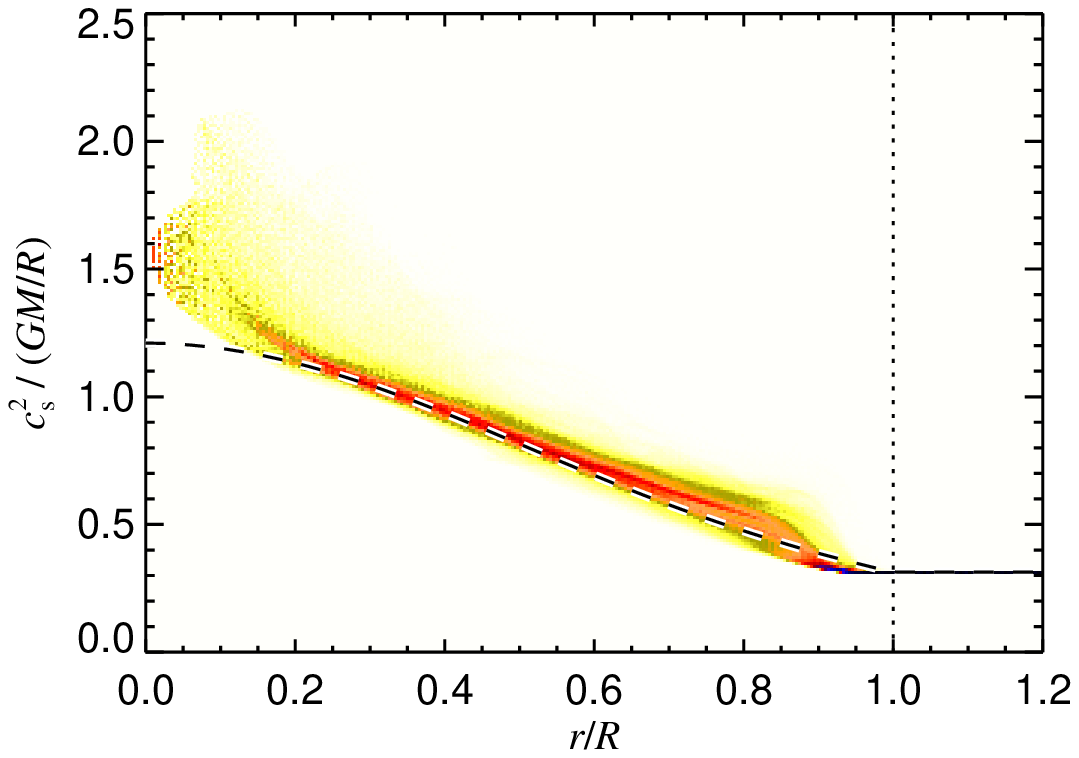}\\
  \plottwo{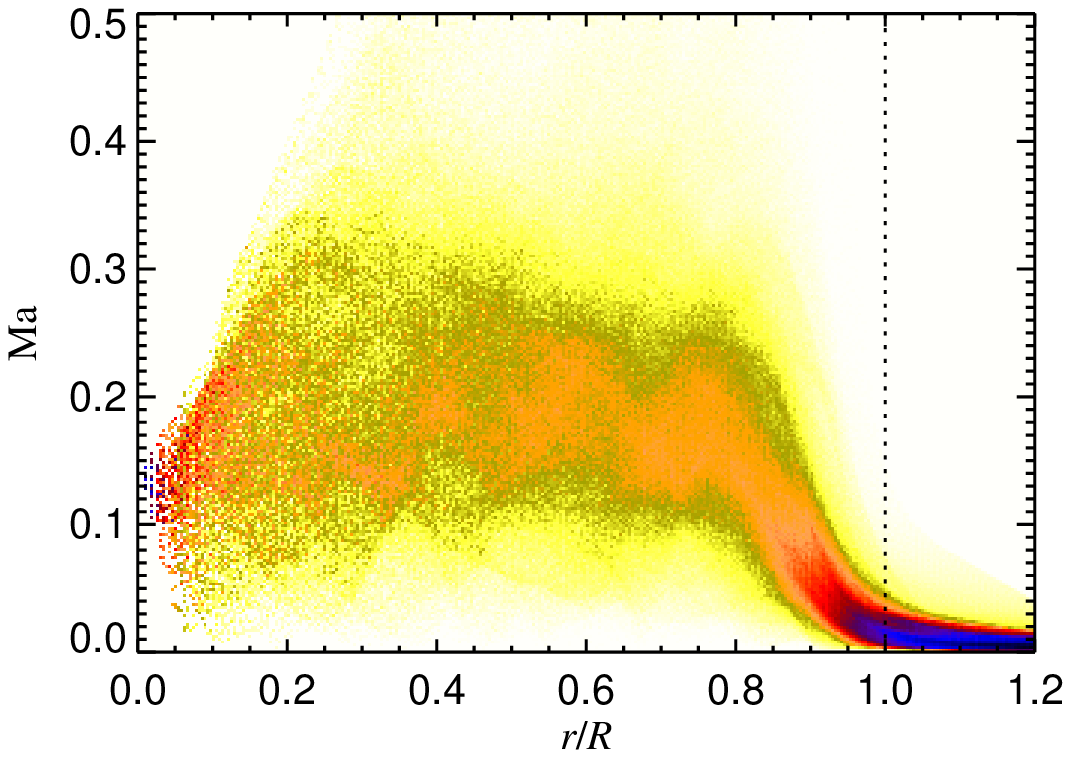}{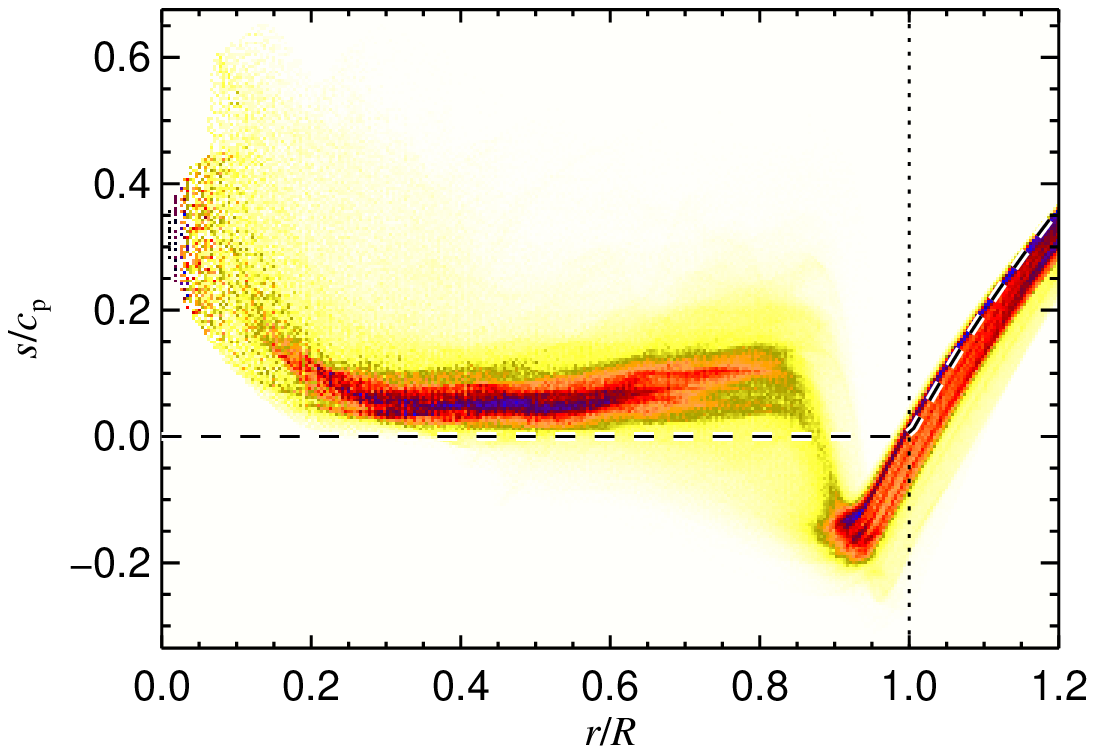}
  \caption{``Histograms'' of
    density $\varrho$,
    squared sound speed $\cs^2$,
    Mach number $\mathrm{Ma}=|\uv|/\cs$,
    and specific entropy $s$
    as a function of radius for the saturated state of Run~1c.
    Dark shades indicate a high probability of the corresponding
    value at a given radius.
    The dashed lines represent the initial profiles.
  }
  \label{Fig-rho-cs-s-r}
\end{figure*}

\subsection{Radial stratification}

Figure~\ref{Fig-rho-cs-s-r} shows density, squared sound speed
(proportional to temperature), Mach number and specific entropy as
function of radius for Run~1c.
Density and squared sound speed vary by
a factor of about $5$ from the center to the stellar surface.
Apart from a few localized transients, the maximum Mach number is below
unity and there is no evidence for shocks.
The total variation in specific entropy is about $0.6\,c_p$.
Even in the bulk of the convection zone ($0.15 < r/R < 0.85$)
the specific entropy has a standard deviation of
about $0.05\,c_p$, which is still much larger than what mixing-length
theory predicts for this type of star.
This is related to the high value of $\mathcal{L}$
that we are using, which is also the reason for the enhanced entropy values
in the core.
The location of the specific entropy minimum is at $r/R\approx0.93$, i.e.\
somewhat below the nominal surface of the star.
This is primarily a consequence of the rather large width of the
profile functions for cooling and velocity damping, which affect the
interior already inside $r=R$.
At that effective radius, we naturally get a thin overshoot layer (as
found in real stellar chromospheres).

\subsection{Hydrodynamic flow patterns}

\begin{figure}[tp]
  \plotone{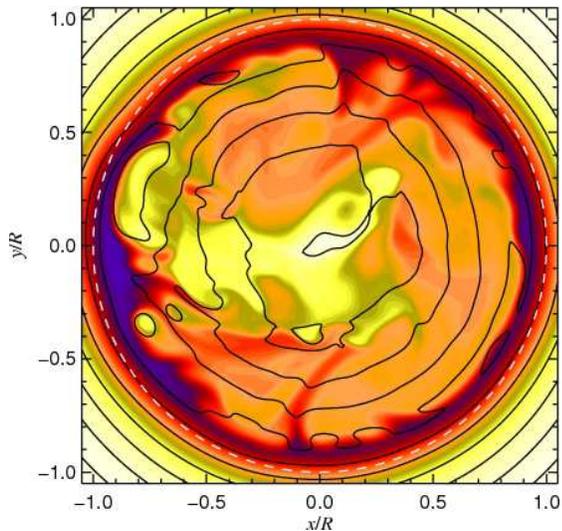}
  \caption{Equatorial section of entropy (color coded: dark for low,
    bright for high entropy) and density
    (isolines) during the kinematic stage of Run~1b (at $t=300\,[t]$),
    where the magnetic field does not affect the dynamics.
  }
  \label{Fig-pslice_5a_256}
\end{figure}

\begin{figure}[t!]
  \plotone{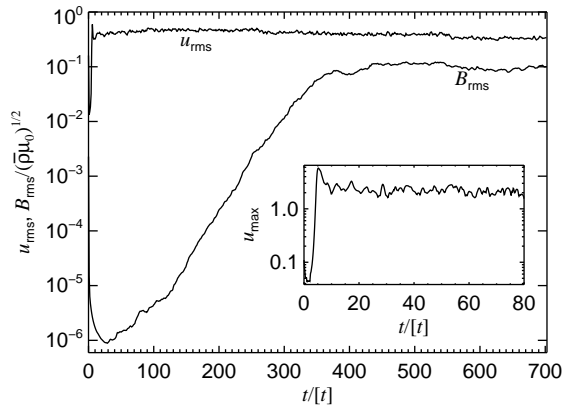}
  \caption{Evolution of root-mean-square velocity $u_{\rm rms}$ and
    magnetic field $B_{\rm rms}$ (represented as an Alfv{\'e}n speed using
    $\overline{\varrho}=0.4\,[\varrho]$ to make the two curves comparable)
    for Run~1b.
    Time is measured in units $[t]$ and velocities in units $[u]$ (see
    Sec.~\ref{S-dimensions}).
    The inset shows the maximum velocity $u_{\rm max}(t)$ during the onset
    of convection --- note the irregular time-behavior.
  }
  \label{Fig-u-B-t-128}
\end{figure}

Since the initial magnetic field is weak (several orders of magnitude
below saturation), the kinematic phase of the dynamo represents the
hydrodynamic flow pattern in a nonmagnetic scenario.
Figure~\ref{Fig-pslice_5a_256} shows an equatorial section of entropy and
density for Run~1b.
One can clearly distinguish narrow cool structures (downdrafts) that are
familiar from box simulations of compressible convection
\citep[e.g.][]{HurlburtEtal:NonlinCompressibleConvection,%
  NordlundEtal:OvershootDynamo}
The flows are far from being laminar, as can also be seen
in the inset to Fig.~\ref{Fig-u-B-t-128}.
However, given the numerical resolution, only a limited range of scales
can be resolved, as can be seen from the magnetic energy spectrum during
the kinematic dynamo phase (see next section).

\begin{figure}[tp]
  \epsscale{0.5}
  \plotone{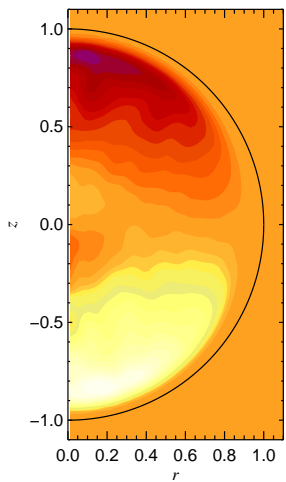}
  \caption{Average of kinetic helicity for Run~2c during the kinematic phase.
    Shown is the azimuthal average of $\uv\cdot\nnabla\times\uv$, averaged
    in time from $t=100\mbox{--}300\,[t]$.
  \label{Fig-helicity-6a_128}
  }
\end{figure}

Figure~\ref{Fig-helicity-6a_128} shows a $t$-$\varphi$ average of kinetic
helicity $\uv\cdot\nnabla\times\uv$ for the kinematic dynamo phase.
As expected from the action of the Coriolis force on expanding upflows and
contracting downflows, the helicity is predominantly negative in the
northern and positive in the southern hemisphere.
If kinetic helicity is connected to a turbulent electromotive force, we
find a distribution of the $\alpha$ effect that is reminiscent of
classical mean-field dynamo models
\cite[e.g.][]{Roberts:KinematicDynamos}.
It should hence not be surprising if the flow generates a large-scale
magnetic field.

\subsection{Dynamo action}

The turbulent kinetic energy quickly reaches a statistically
steady state after about 5 dynamical times
($t\approx5\,[t]$), while the energy of the initially random
magnetic field decays at first; see Fig.~\ref{Fig-u-B-t-128}.
This is because most of the magnetic energy of the random field
is in the small scales and thus gets quickly dissipated.
The magnetic field then grows exponentially with a growth rate
$\lambda=\dd \ln B_{\rm rms}/\dd t$ of about $0.04/[t]$ (for Run~1b).

\begin{figure}[t!]
  \epsscale{1}
  \plotone{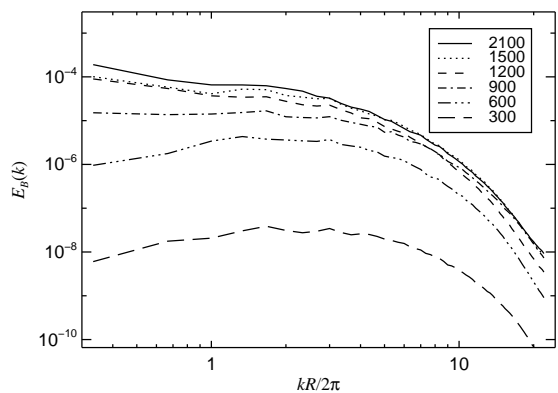}
  \caption{Spectra of the magnetic field for times
    $t=(300,600,900,1200,1500,2100)\,[t]$ of Run~2c (for this run, exponential
    field growth levels off around $t \approx 700\,[t]$).
    Magnetic energy increases with time and eventually reaches saturation.
    At late times the largest scales dominate.
  }
  \label{Fig-spectra-128}
\end{figure}

\begin{figure}[t!]
  \plotone{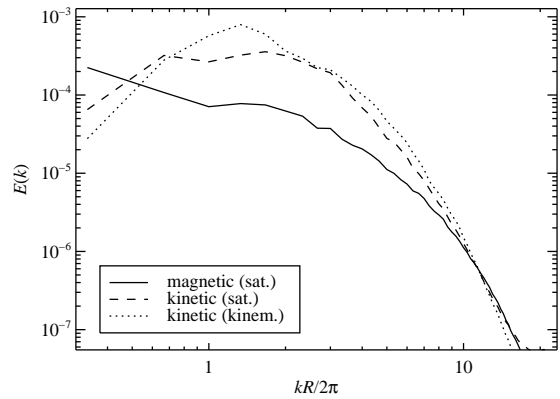}
  \caption{Magnetic (solid line) and kinetic (dashed line) power spectra
    for the saturated phase of Run~2c. 
    While the velocity spectrum peaks around $k\approx 2\pi/R$, magnetic
    energy is distributed more flatly around the largest scales.
    For comparison, the dotted line shows the kinetic spectrum during the
    kinematic dynamo phase.
  }
  \label{Fig-spectra-128-kin}
\end{figure}

During the kinematic stage of the dynamo, the magnetic field grows
exponentially with the same rate at all wavenumbers,
so the spectrum remains shape-invariant,
as can be seen in Fig.~\ref{Fig-spectra-128}.
The maximum of the magnetic spectrum is around $k \approx 3\times 2\pi/R$.
The growth time $1/\lambda$ is about one order of magnitude shorter than
the global diffusive time scale $\tau_{\rm Ohm}$, which is a manifestation
of turbulent magnetic diffusion.

At later times, magnetic energy saturates first at the smallest
scales, while the large scales still accumulate energy.
Eventually all scales are saturated, but now the magnetic spectrum peaks
at a larger scale than during the kinematic stage.
As the magnetic field reaches saturation, the kinetic energy of the flow
is decreased by a certain amount that depends on the relative importance
of rotation (see Table~\ref{T-sum-1}).
For slowly rotating spheres the kinetic energy decreases by only about
10--20\% (Runs~1a and 2b), but for more rapidly rotating spheres,
where the magnetic energy is also much larger, the suppression of the
kinetic energy is about 50--60\% (Runs~2c--2e).
The strong dependence of the kinetic energy on the magnetic field strength
suggests that the flows are probably not strongly turbulent and still
governed by a large scale more laminar flow pattern.

Increasing the resolution by a factor of 2, while at the same time
decreasing dissipative effects (cf.~Runs~1a, 1b), we see that
the growth rate increases significantly (by a factor of 2.5, see Table
\ref{T-sum-1}), but in the saturated
state the rms velocity changes insignificantly.
The rms magnetic field increases by about 40\%, which is rather large
and may be a consequence of the dynamo not being strongly supercritical.
Decreasing the luminosity by a factor of 2 (cf.~Runs 1b, 1c)
decreases rms velocity and magnetic field only by about 20\%.

Figure~\ref{Fig-spectra-128-kin} shows spatial spectra of kinetic and
magnetic energy.
Kinetic energy peaks at a wavenumber of about
$k_{\rm p} \approx 1\times 2\pi/R \approx 6\,[x]^{-1}$, which corresponds to the
energy-carrying scale [smaller scales have a negligible contribution to
the total kinetic energy $\int E(k)\dd k$].
The corresponding turnover time is
$\tau = (u_{\rm rms}k_{\rm p})^{-1} \approx 2\,[t]$.
This results in a normalized growth rate $\lambda\tau=0.08$, which
is comparable to the values for both helically and nonhelically forced
turbulence simulations where $\lambda\tau=0.03\mbox{--}0.1$,
see \cite{Brandenburg:InverseCascade} and
\cite{HaugenEtal:NonhelicalTurbulence}, respectively.
The saturation value of magnetic energy is typically an order
of magnitude below the kinetic energy of the turbulence for the slowly
rotating models.
This is quite similar to the ratio found in earlier simulations of
convection driven dynamos in Cartesian and spherical geometries
\citep{MeneguzziPouquet:ConvectionDynamos,%
  NordlundEtal:OvershootDynamo,%
  Brun:Interaction}.
For the faster rotating Runs~2c--2e, however, magnetic energy is
comparable to kinetic energy.

Theoretically, there is always the possibility of different solutions to
a nonlinear problem, depending on the initial conditions.
This possibility has been anticipated in connection with the geodynamo
\citep{RobertsSoward:DynamoTheory}, but it has so far not been seen in any
turbulent dynamo simulation \citep[e.g.][]{GlatzmaierRoberts:NatureI}.
In principle there is even the possibility of so-called self-killing
dynamos that decay after full saturation has been reached, but such behavior
has so far only been found under rather artificial conditions and in
the absence of turbulence \citep{FuchsEtal:Selfkilling}.
In some cases we have restarted our simulations from a snapshot that
has been obtained for different parameters.
In such cases we always recovered statistically the same solution that
was obtained in the standard way by starting from a weak seed magnetic
field and a non-convecting initial state.
This was further confirmed by restarting Run~2c
with a $10^{10}$ times weaker initial field, which led to an indistinguishable
time history of $u_{\rm max}$ for $t < 5\,[t]$,\footnote{
  The resulting Lyapunov time scale is a few turnover times
  $R/u_{\rm rms}$, as one would expect.
}
and to statistically equivalent behavior for larger times.

The saturation appears to happen on a dynamical time scale,
i.e.~we see no evidence for resistively
limited saturation, as it was found in helically forced simulations in
a triply periodic domain \citep{Brandenburg:InverseCascade}.
In Run~1b, the total simulated time $t\approx 700\,[t]$ corresponds to
about $2\,\tau_{\rm Ohm}$.
The nonresistive saturation behavior could be due to the fact that in
the present simulations the boundaries are open and permit a magnetic
helicity flux across the equatorial plane and out of the box
\citep{BrandenburgDobler:Boundaries,Brandenburg:OpenClosed}.
Another possibility is that the magnetic Reynolds number is still too
small for magnetic helicity conservation to have an effect.

\subsection{Dependence on rotation rate}
In Runs~2a--2e, we vary the rotation rate $\Omega_0$, while keeping all
other parameters fixed.
As the rotation rate is increased, the root-mean-square velocity of
the turbulence decreases.
This is to be expected, because the presence of rotation is known to
delay the onset of convection
\citep[e.g.][]{Chandrasekhar:HydroStability}.
The rms magnetic field strength increases monotonically with the Coriolis
number (or the rotation rate $\Omega_0$) for the runs shown.

In the absence of rotation (Run~2a), there is no net helicity or net shear
and hence no reason for the generation of a {\it large-scale}
magnetic field.
However, we still find that the magnetic energy increases, albeit more slowly
and to a lower value than for the rotating runs.
This is a manifestation of the `fluctuating' or `small-scale' dynamo 
\citep{Kazantsev:Enhancement,%
  MeneguzziEtal:HelicalNonhelical,%
  Cattaneo:SmallScaleynamo},
which requires a considerably larger value of the magnetic Reynolds number
than the helical dynamo, and there are indications that Run~2a is
only mildly supercritical.

\begin{figure}[t!]
  \plotone{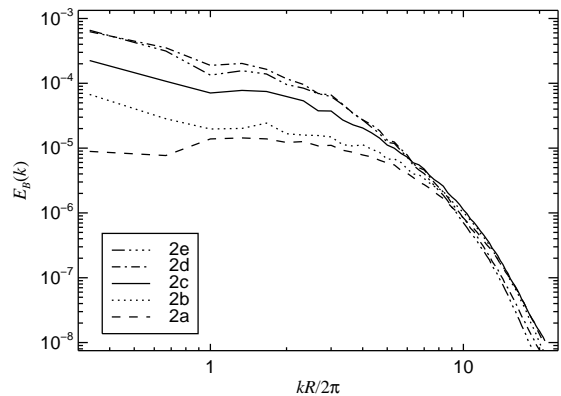}
  \caption{Magnetic energy spectra for the saturated states of
    Runs~2a, 2b, 2c, 2d, and 2e.
  }
  \label{Fig-spectra-6a-f}
\end{figure}

Comparing the magnetic energy spectra for runs with different rotation
rates (Fig.~\ref{Fig-spectra-6a-f}), we find that the magnetic energy at
the large scales increases with $\Omega_0$ at least up to
$\Omega_0 \approx 5$ (corresponding to $\mbox{Co} \approx 100$),
while the small scales are only weakly affected by rotation.
The saturation for rapid rotation has been predicted by mean-field dynamo
theory,
\citep{RuedigerKitchatinov:AlphaQuenching,%
  OssendrijverEtal:MagnetoconvectionI},
and for even larger $\Omega_0$, one expects a reduction of large-scale
dynamo efficiency.
However, for our models the peak dynamo efficiency occurs for rather large
values of $\mbox{Co}$.
Another surprise is that dynamo activity at small scales
($kR/2\pi \gtrsim 10$) is not quenched for `superfast' rotation, although
the Coriolis force should play a significant role until $k$ reaches
significantly larger values.

\begin{figure}[t!]
  \plotone{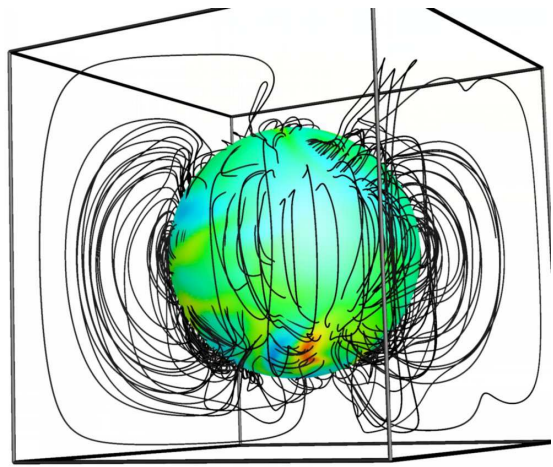}
  \caption{Three-dimensional visualization of the magnetic field for
    Run~2c at $t=2600\,[t]$ (saturated phase).
    Magnetic field lines are shown, together with the surface of the
    sphere.
  }
  \label{Fig-flines-128-sat}
\end{figure}

\begin{figure}[t!]
  \plottwo{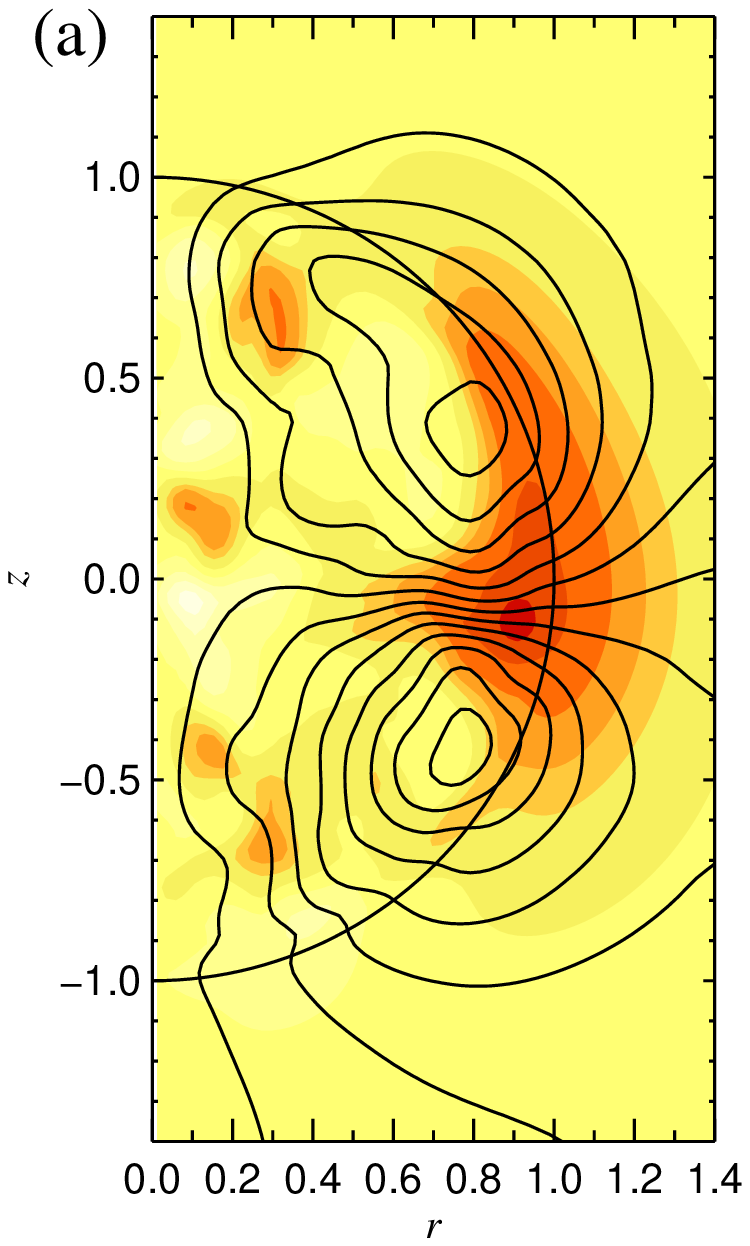}{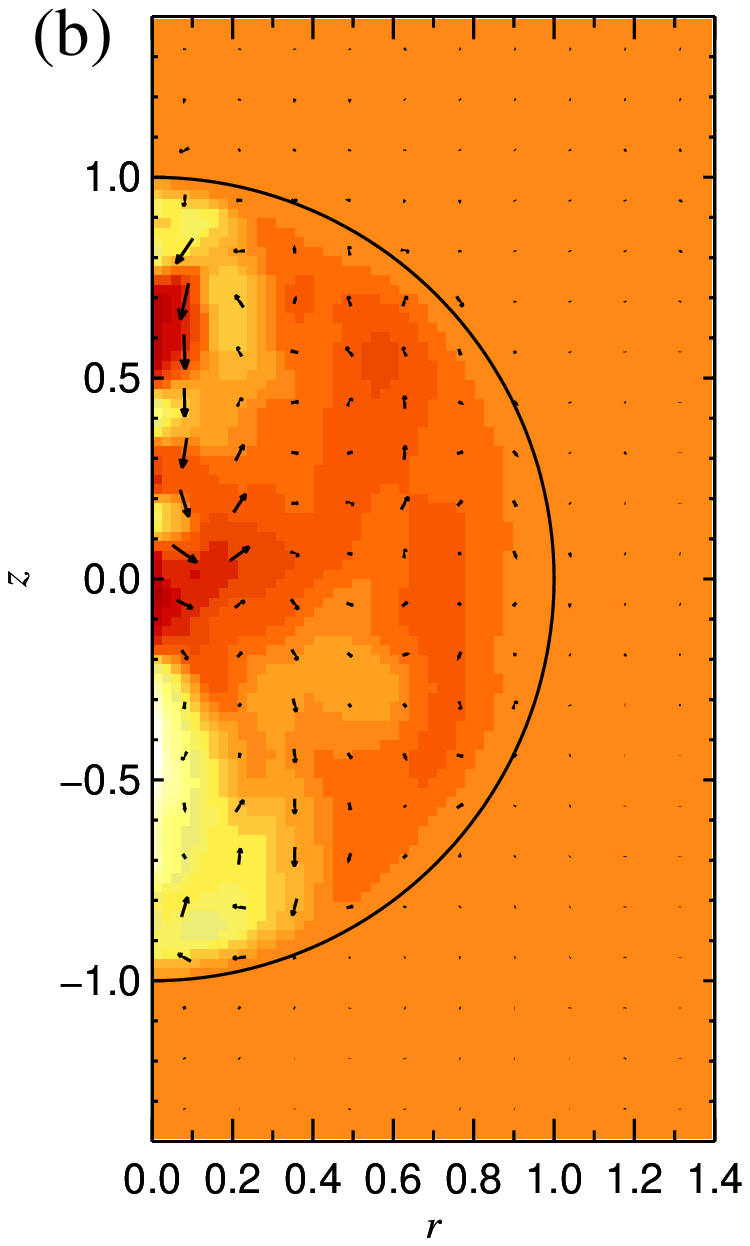}
  \caption{Azimuthal averages of magnetic field (a) and velocity (b)
    for Run~2c at time $t=2100\,[t]$ (saturated phase).
    Panel (a) shows poloidal field lines of the
    $\varphi$-averaged magnetic field
    (orientation of the field lines in the top half is predominantly
    counter-clockwise), superimposed on a color/gray-scale representation
    of the azimuthal mean field,
    $\overline{B_\varphi}$, with bright colors representing
    $\overline{B_\varphi}>0$ and dark colors representing
    $\overline{B_\varphi}<0$.
    Note the mixed parity of the structure of the mean field with a strong
    quadrupolar contribution.
    Panel (b) shows vectors of the mean poloidal velocity superimposed
    on a color/gray scale representation of the mean angular velocity
    $\delta\overline{\Omega}(r,z)\equiv\overline{u_\varphi}/(r\sin\theta)$
    with bright colors for $\delta\overline{\Omega}>0$
    and dark colors for $\delta\overline{\Omega}<0$.
    Note that for different snapshots in time the velocity field
    $\overline{\uv}$ looks very different.
  }
  \label{Fig-phiaverages-128-sat}
\end{figure}

\begin{figure*}
  \plotone{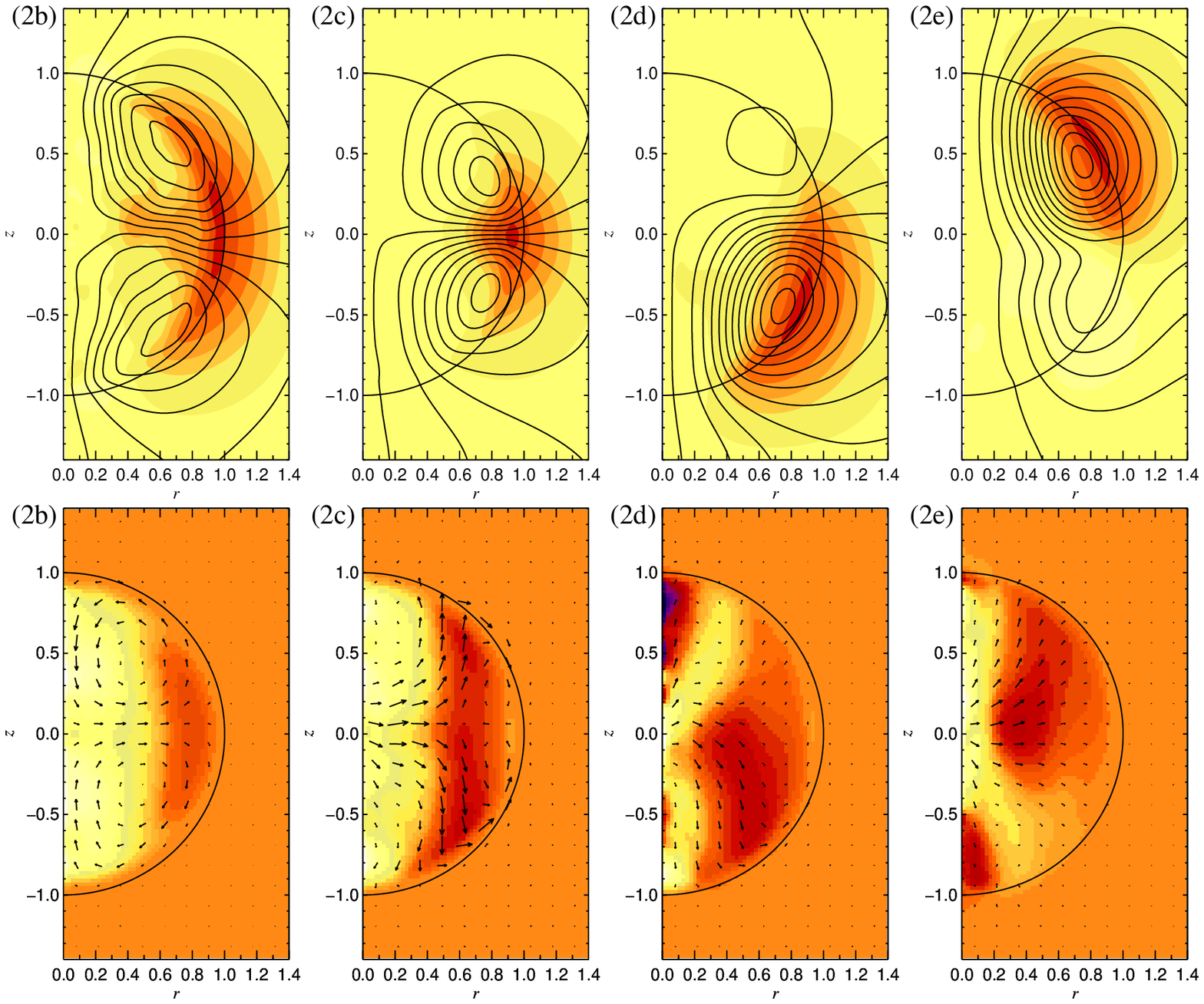}
  \caption{
    Azimuthal averages for Runs~2b--2e as in
    Fig.~\ref{Fig-phiaverages-128-sat}, but now
    the fields are additionally averaged over $1000$--$2000$ time units
    [$t$] (about $150$--$300$ turbulent turnover times, or
    $4$--$8$ $\tau_{\rm Ohm}$) during the saturated state.
    Angular velocity $\Omega_0$ increases from left to right;
    the panels are labeled by the names of the individual runs.
    Top row: magnetic field.
    Orientation of the field lines is predominantly counterclockwise in
    the top hemisphere and clockwise in the bottom hemisphere in all four
    panels.
    Bottom row: velocity.
    Note that the angular velocity shows some tendency to be constant along
    cylinders for Runs~2b and 2c, while magnetic and velocity field are
    asymmetric for Runs~2d and 2e.
    The amplitudes of magnetic and velocity fields have been scaled
    individually for each image; absolute values of $B_{\rm rms}$ and
    $u_{\rm rms}$ are given in Table~\ref{T-sum-1}.
  }
  \label{Fig-phiaverages-tavg-128-sat}
\end{figure*}

\subsection{Large-scale field structure}

Although a lot of the magnetic energy is due to the small-scale structures,
as can be seen in the magnetic energy spectra, outside the star the field
shows features of a dipole-like structure with a noticeable contribution from
the first few multipoles; see Fig.~\ref{Fig-flines-128-sat} where we
show a visualization of the three-dimensional magnetic field lines.

A more quantitative presentation of the large-scale magnetic and velocity
fields is obtained by considering azimuthal averages,
as shown in Fig.~\ref{Fig-phiaverages-128-sat} for one snapshot of Run~2c.
The magnetic field shows a clear large-scale component with predominantly
quadrupolar symmetry with respect to the midplane, but still including
dipolar contributions.
The velocity field shows little large-scale structure and varies strongly
in time.

We find a considerably more regular structure when applying time averaging
to the azimuthally averaged data.
In Fig.~\ref{Fig-phiaverages-tavg-128-sat} we show, for the saturated
states, the correspondingly averaged magnetic fields (top row) and
velocity fields (bottom row) for four different runs with rotation rate
$\Omega_0$ increasing from left to right.
We note that for all runs the averaged velocity field changes very little
from the kinematic to the saturated stage of the dynamo.

\begin{deluxetable*}{crllllllll} 
  \tablecaption{Additional diagnostic quantities for Runs~2a--2e.
  \label{T-sum-2}
  }
  \tablewidth{0.65\textwidth}
  \tablehead{
    \colhead{Run}
        & \colhead{$\Omega_0$}
                 & \colhead{$q_{\mathbf{u}}$}
                          & \colhead{$q_{\mathbf{B}}$}
                                   & \colhead{$p_{\mathbf{u}}$}
                                            & \colhead{$p_{\mathbf{B}}$}
                                                     & \colhead{$\Delta\omega^{\rm kin\,/\,sat}$}
                                                                         & \colhead{$\sigma_{\mathbf{u}}^{\rm kin\,/\,sat}$}
                                                                                               & \colhead{$\sigma_{\mathbf{B}}^{\rm kin\,/\,sat}$}
                                                                                                                     & \colhead{$P$}
 }
  \startdata
    2a  & $0.0$  & $0.30$ & $0.02$ & $0.04$ & $1.10$ & $0.53$\,/\,$0.24$ & $0.240$\,/\,$0.214$ & $0.071$\,/\,$0.079$ & $-0.12$ \\
    2b  & $0.5$  & $0.38$ & $0.19$ & $1.44$ & $2.95$ & $1.23$\,/\,$0.97$ & $0.222$\,/\,$0.221$ & $0.077$\,/\,$0.147$ & $+0.99$ \\
    2c  & $2.0$  & $0.14$ & $0.39$ & $1.80$ & $2.36$ & $0.56$\,/\,$0.22$ & $0.062$\,/\,$0.079$ & $0.124$\,/\,$0.271$ & $+0.91$ \\
    2d  & $5.0$  & $0.14$ & $0.40$ & $1.59$ & $1.32$ & $0.82$\,/\,$0.26$ & $0.054$\,/\,$0.064$ & $0.046$\,/\,$0.271$ & $+0.28$ \\
    2e  & $10.0$ & $0.14$ & $0.46$ & $1.27$ & $1.17$ & $0.50$/\,$\,0.14$ & $0.080$\,/\,$0.063$ & $0.047$\,/\,$0.326$ & $-0.28$ \\
  \enddata%
  \tablecomments{
    All quantities refer to the saturated state, unless explicitly
    labeled with `kin'.
  }
\end{deluxetable*} 

With this averaging, we find almost perfect quadrupolar symmetry for
$\mathbf{B}$ in Runs~2b and (slightly less pronounced) 2c.
On the other hand, Runs~2d and 2e show very pronounced hemispheric
asymmetry that appears to be relatively long-lived.
For Runs~2b and 2c, the velocity field shows a meridional circulation
pattern that is directed outwards at the equator, and surfaces of constant
angular velocity $\omega$ are approximately cylindrical.
For the rapidly spinning Runs~2d and 2e, the asymmetry in the magnetic
structure is reflected in differential rotation and meridional circulation.

An explicit measure for the efficiency of large-scale field generation
is the ratio
\begin{equation}
  q_{\mathbf{B}}
  \equiv \frac{\bigl< \bigl< \overline{\mathbf{B}} \bigr>_t^{\,2} \bigr>_{rz}^{\,1/2}}
              {B_{\rm rms}} \; ,
\end{equation}
which is given in Table~\ref{T-sum-2}.
A similar quantity $q_{\mathbf{u}}$ is also defined for the velocity.
Here the overbar denotes azimuthal averaging, $\left<\cdot\right>_t$
represents time averaging, while $\left<\cdot\right>_{rz}$ denotes spatial
averaging over the sphere, and
$B_{\rm rms} = \langle\langle\overline{\mathbf{B}^2}\rangle_t\rangle_{rz}$.
This ratio is $q_{\mathbf{B}}=0.19$ for Run~2b and increases further with
the rotation rate.
These values are quite large, suggesting that large-scale field generation
is quite efficient.
However, in forced turbulence simulations with open boundaries and no
shear \citep{BrandenburgDobler:Boundaries},
$q_{\mathbf{B}}$ decreases with
increasing magnetic Reynolds number.
On the other hand, simulations of forced turbulence suggest that the presence of
shear is critical for allowing the dynamo amplitude to be independent
of the magnetic Reynolds number \citep{Brandenburg:OpenClosed}.
Further numerical simulations are necessary to see whether the same
behavior occurs here as well.

The ratios
\begin{equation}
  p_{\mathbf{u}}
  \equiv \frac{\bigl< \bigl< \overline{u_\varphi} \bigr>_t^{\,2}
               \bigr>_{rz}^{\,1/2}}
              {\bigl< \bigl< \overline{u_{\rm pol}} \bigr>_t^{\,2}
               \bigr>_{rz}^{\,1/2}} \; , \qquad
  p_{\mathbf{B}}
  \equiv \frac{\bigl< \bigl< \overline{B_\varphi} \bigr>_t^{\,2}
               \bigr>_{rz}^{\,1/2}}
              {\bigl< \bigl< \overline{B_{\rm pol}} \bigr>_t^{\,2}
               \bigr>_{rz}^{\,1/2}}
\end{equation}
quantify the importance of the azimuthal components in the $\varphi$- and
$t$-averaged fields.
In the non-rotating case, $p_{\mathbf{u}}$ is very small, indicating that
systematic azimuthal flows are weak.
With increasing angular velocity, however, $p_{\mathbf{u}}$ at first
increases as systematic differential rotation evolves.
The slight decline of $p_{\mathbf{u}}$ in Runs~2d and 2e is connected with
the stronger magnetic field in those cases.
In fact, both the azimuthal component
$\langle \langle \overline{u_\varphi} \rangle_t^{\,2}\rangle_{rz}^{\,1/2}$ and
the poloidal component
$\langle \langle \overline{u_{\rm pol}} \rangle_t^{\,2} \rangle_{rz}^{\,1/2}$
decrease monotonically from Runs~2b to 2e because of the increasing
large-scale magnetic field.

The ratio $p_{\mathbf{B}}$ is 1.1
in the nonrotating case, it has a maximum already for $\Omega_0=0.5$
(Run~2b) and then declines.
The total magnetic energy continues to grow with higher rotation rates
(see Table~\ref{T-sum-1}),
indicating that with increasing rotation rate the poloidal field
continues to grow, while the toroidal field remains roughly unchanged.
This seems to be in qualitative agreement with the simulations by
\cite{BrunEtal:GlobalScale} for dynamos in convective shells without
tachocline.
On the other hand, this ratio is in all cases small compared to what is
expected for the Sun, where the tachocline can be expected to have a
strong effect.

With the exception of Run~2b, the absolute amplitude $\Delta\omega$ of
the differential rotation for kinematic and saturated states is similar.
This is consistent both with theory \citep{KitchatinovRuediger:DiffRot}
and with observations showing only a weak dependence of surface
differential rotation on stellar rotation for late type stars
\citep{BarnesEtal:DependenceRotation}, which are however not fully
convective.

Next, we show the energy ratios
\begin{equation}
  \sigma_{\mathbf{u}}
  \equiv \frac{\langle\langle\overline{\mathbf{u}}^2\rangle_t
               \rangle_{rz}}
              {u_{\rm rms}^2} \; ,
  \qquad
  \sigma_{\mathbf{B}}
  \equiv \frac{\langle\langle\overline{\mathbf{B}}^2\rangle_t
               \rangle_{rz}}
              {B_{\rm rms}^2} \; ,
\end{equation}
quantifying the fraction of energy contained in the axisymmetric part of
$\mathbf{u}$ and $\mathbf{B}$.
For most of the rotating runs, $\sigma_\mathbf{B}$ increases drastically
from the kinematic stage to saturation.
This is another manifestation of the trend towards large-scale fields once
the small scales are saturated (see Fig.~\ref{Fig-spectra-128}).
On the other hand, $\sigma_\mathbf{u}$, which is strongly reduced by
rotation, is not severely affected by the magnetic field saturation.

Finally, we consider the parity of the mean field with respect to the
equatorial plane.
Earlier work on mean-field dynamos in full spheres
\citep{BrandenburgEtal:Parity} has shown that for weakly supercritical
dynamos the parity is dipolar (antisymmetry with respect to the equator).
However, as the dynamo becomes more supercritical, the parity can become
quadrupolar (symmetric with respect to the equator), but mixed and chaotic
behaviors are also possible \citep{CovasEtal:Geometry}.
Parity of the mean field can be quantified as
\begin{equation}
  P=\frac{E_{\rm S}-E_{\rm A}}
         {E_{\rm S}+E_{\rm A}} \; ,
\end{equation}
where $E_{\rm S}$ and $E_{\rm A}$ denote the energies contained in the
symmetric and antisymmetric parts of $\overline{\mathbf{B}}$.
The values of $P$ are listed in Table~\ref{T-sum-2} for the saturated
stage of the dynamo.
For Runs~2b and 2c the mean field is nearly
quadrupolar $P\approx+1$, as is also evident from
Fig.~\ref{Fig-phiaverages-tavg-128-sat}.\footnote{%
  Note that the mean field for the snapshot shown in
  Fig.~\ref{Fig-flines-128-sat} is still predominantly symmetric, although
  its poloidal field lines look quite dipolar.
  This is due to three factors: the dominance of the azimuthal component,
  the non-axisymmetry of the field, and a hemispheric asymmetry that is
  not very prominent in Fig.~\ref{Fig-flines-128-sat}.
}
Both in the absence of rotation and for strong rotation the mean fields
are of more mixed parity character.
Comparison with mean-field dynamos would suggest that our Runs~2b and
2c are in the ``more supercritical'' regime.
However, we have not found a case that would clearly belong to the
weakly supercritical regime, where dipolar fields are expected.
In addition, mean-field theory would suggest cyclic mean fields, that
have also not been seen.
It is possible that such features would emerge in a direct simulation
only after many more turnover times than what has been possible here.

The ratios $q$ and $p$ allow us primarily to assess the mode of operation
of the dynamo found in the simulations.
In particular, they provide a sensitive tool to assess the possible
dependence of the magnitude of large-scale fields on the magnetic
Reynolds number.
Observationally, of course, only the limit of very large magnetic
Reynolds numbers is relevant.
More detailed comparison of the $q$ and $p$ ratios with observations
is hampered by the fact that the magnetic field in the star's interior
cannot be measured with current techniques.
The interpretation of proxies such as filling factors and the appearance
of magnetic fields at the stellar surface may be premature as long as
we do not fully understand the connection between magnetic fields at the
surface and the interior.
For example, the interpretation of bipolar spots in terms of distinct
flux tubes may not be valid and hence the presence of spots may not
necessarily indicate a high degree of intermittency in the interior
\citep{Brandenburg:OpenClosed}.
However, once the physics of the stellar surface is modeled more
realistically (e.g., without imposing an artificial cooling layer to
model radiation) it would be useful to produce synthetic surface maps
and light curves that can be compared with observations.

\section{Conclusions}

The present work suggests that fully convective stars are capable
of generating not only turbulent magnetic fields, but also strong
large-scale fields that dominate the magnetic energy spectrum.
In most of our models, the large-scale field has a strong quadrupolar
component, in contrast to what is
expected from mean-field theory for dynamo action in thick shells
and in full spheres
\citep{Roberts:KinematicDynamos}.
We have so far not seen evidence of magnetic cycles.
The resolution of our models is still too low to be able
to tell whether this type of magnetic field generation will continue to
operate
at much larger magnetic Reynolds numbers, but our results disprove the
claim that a strong shear layer or a stably stratified core are necessary
ingredients for the generation of large-scale magnetic fields.
As one would expect in the absence of strong shear layers, the toroidal
and poloidal components of the mean magnetic field are roughly comparable.

Another important result concerns the self-consis\-tently produced
differential rotation.
In our simulations, the angular velocity shows some tendency to
be constant along cylinders, which is plausible for rapidly rotating
stars.
Whether or not this is realistic is difficult to say.
Asteroseismology may in the future be able to reveal the internal angular
velocity of stars, but at present the time coverage is still too short
and incomplete.
There is at least some hope of observing the surface differential rotation,
at least of sufficiently rapidly rotating fully convective stars such as
T Tauri stars, using surface imaging
\citep{CollierCameronEtal:Astrotomography}.
This would be particularly interesting, given that our simulations
predict a more slowly rotating equator.
This behavior is opposite to that in the solar case.
Thus far, theory in terms of the $\Lambda$ effect
\citep[e.g.][]{RuedigerHollerbach:MagneticUniverse}
also tends to
produce a faster equator, unless the turbulent motions possess a
predominantly radial structure ($u_{r,{\rm rms}} \gg u_{\varphi,{\rm rms}}$).
In our case, however, there is strong meridional circulation, which,
due to conservation of angular momentum,
causes the outer layers to rotate more slowly;
see \cite{KitchatinovRuediger:AntiSolar}.
Again, this result may no longer hold in real stars, because
in our model the degree of stratification is far too low and the luminosity
too high, so the convective velocities and meridional circulation
tend to be exaggerated.

Another reason for the slowly rotating equator could be connected with
the outer boundary condition.
In connection with geodynamo simulations there are indications that a no-slip
outer boundary condition (with respect to a rigidly rotating sphere) tends to
produce a more slowly rotating equator
\citep{ChristensenEtal:ParameterStudy}.
In the limit of a short damping time, our effective outer boundary
condition at $r=R$ should
indeed be closer to a no-slip condition than to a free-slip condition.
On the other hand, in strongly magnetized stars the coronal magnetic field
may enforce a rigidly rotating exterior, and hence produce conditions
close to what is represented by our model.

\acknowledgments
We thank Matthew Browning and Juri Toomre
for detailed comments on a draft of our paper
and an anonymous referee for suggesting many improvements to the paper.
We acknowledge support from the Isaac Newton Institute in Cambridge,
where part of this work has been completed.
The Danish Center for Scientific Computing is acknowledged
for granting time on the Linux cluster in Odense (Horseshoe).


\appendix

\section{Reference model}
\label{RefModel}

In order to specify the initial conditions and the gravity potential, we
use a simple spherically symmetric, hydrostatic, self-gravitating,
isentropic model.
The equations for this reference model are
\begin{eqnarray}
  \label{1d-mass}
  \frac{dm_r}{dr}       &=& 4\pi r^2 \varrho \; , \\
  \label{1d-density}
  \frac{d\varrho}{dr}   &=& -\frac{G\,m_r\varrho^{2-\gamma}}
                                  {\gamma K r^2} \; ,
\end{eqnarray}
where $m_r$ denotes the total mass inside the sphere of radius $r$,
together with the boundary conditions
\begin{equation}
  m_r(0) = 0 \; , \qquad
  \varrho(0) = \varrho_{\rm c} \; .
\end{equation}
Here $K=e^{\gamma s_0/c_p}=\const$ is the polytropic constant, relating
pressure $p$ and density $\varrho$ via $p=K \varrho^\gamma$, and $s_0$ is
the constant value of entropy.
The adiabatic exponent is $\gamma=5/3$, and thus our reference model is a
polytropic model with polytropic index $m=3/2$. 

Equations (\ref{1d-mass}) and (\ref{1d-density}) are integrated
outwards, starting with certain values
$(\varrho_{\rm c}, s_0)$ for central density and entropy.
As is common with polytropic models, the solution can have a surface
(where $\varrho=p=T=0$) at some finite radius $R_{\rm surf}$, which must
not be smaller than the desired stellar radius $R$.

Varying the central values $(\varrho_{\rm c}, s_0)$, we can tune the
reference model to match a given reference stellar radius $R$ and total
mass $M$.
We choose the values
$R = 0.27\,R_\odot = \unit[1.9\EE{8}]{m}$, and
$M = 0.21\,M_\odot = \unit[4.2\EE{29}]{kg}$,
which correspond to an M5 dwarf with a luminosity of
$L = 0.008\,L_\odot = \unit[3\EE{24}]{W}$.

Once the temperature and density profiles are known, one can calculate the
approximate volume heating rate according to the
formula
\begin{equation} \label{Heat-rho-Temp}
  \Heat(\varrho,T)
  \approx4.9\times10^{-4}
    \left(\frac{\varrho}{\unit[10^5]{kg\,m^{-3}}}\right)^2
    \left(\frac{T}{10^6\,\mathrm{K}}\right)^{5.3} {\rm W\,m^{-3}} \; .
\end{equation}
The $T^{5.3}$ dependence is an approximation for the $pp1$ chain of
hydrogen burning near
$T_{\rm c}\approx 6\EE{6}\,\mathrm{K}$;
see Sec.~18.5.1 of
\cite{KippenhahnWeigert:StellarStructureEvolution}.
We have used a Gaussian approximation to this (time-independent) radial
dependence of $\Heat(r)$ for all simulations presented here, while
adjusting the total luminosity by a multiplicative factor.
Figure~\ref{Fig-Lr-r} shows the luminosity
$L_r(r)$ as a function of radius.

\section{High-order upwind derivatives}
\label{S-upwind}
\newcommand{\Order}[1]{O\left(#1\right)}

Convection simulations with high-order centered finite-difference schemes
sometimes show a tendency to develop `wiggles' (Nyquist zigzag) in $\ln\varrho$.
This can be avoided by using a high-order upwind derivative operator, where the
point furthest downstream is excluded from the stencil. 
We apply this technique only to the terms $\uv\cdot\nnabla\ln\varrho$ and
$\uv\cdot\nnabla s$.
In the following we discuss the treatment in the $x$ direction, but the
treatment for the other directions is analogous.
For $u_x>0$, we replace
\begin{equation}
  D_{\rm cent,6}\, f_0
  = \frac{-f_{-3} + 9 f_{-2} - 45 f_{-1} 
           + 45 f_{1} - 9 f_{2} + f_{3}}
         {60\,\delta x}
  = f'_0 + \frac{\delta x^6\,f^{(7)}(\xi_6)}{140} \; ,
  \qquad x_{-3} < \xi_6 < x_3 \; ,
\end{equation}
by
\begin{equation}
  D_{\rm up,5}\, f_0
  = \frac{-2 f_{-3} + 15 f_{-2} - 60 f_{-1} + 20 f_{0} + 30 f_{1} - 3 f_{2}}
         {60\,\delta x}
  =f'_0 - \frac{\delta x^5\,f^{(6)}(\xi_5)}{60} \; ,
  \qquad x_{-3} < \xi_5 < x_2 \; .
\end{equation}
Both formul\ae{} follow from Markoff's formula
\citep[\S 25.3.7]{AbramowitzStegun}.

The difference between the sixth-order central and fifth-order upwind
derivative is proportional to the sixth derivative operator
\begin{equation}
  D^6_{\rm cent,2}\, f_0
  = \frac{f_{-3} - 6 f_{-2} + 15 f_{-1} - 20 f_{0}
          + 15 f_{1} - 6 f_{2} + f_{3}}
         {\delta x^6}
  =  f^{(6)}_0 + \frac{\delta x^2 f^{(8)}(x_0)}{4}
     + O(\delta x^4)\; ,
\end{equation}
namely
\begin{equation}
  D_{\rm up,5}\, f_0
  = D_{\rm cent,6}\, f_0 - \alpha\, \delta x^5 D^6_{\rm cent,2} \; ,
\end{equation}
with $\alpha=1/60$.
This allows us to represent the fifth-order upwind scheme in the advection
term (for both signs of $u_x$) by sixth-order hyper-diffusion:
\begin{equation}
  -u_x f'_{\rm up,5}
  = -u_x f'_{\rm cent,6}  + \alpha |u_x|\,\delta x^5 f^{(6)}_{\rm cent,2} \; .
\end{equation}


\bibliographystyle{apj}
\bibliography{paper}


\end{document}